\documentclass[graybox]{svmult}

\usepackage{mathptmx}       % selects Times Roman as basic font
\usepackage{helvet}         % selects Helvetica as sans-serif font
\usepackage{courier}        % selects Courier as typewriter font
\usepackage{type1cm}        % activate if the above 3 fonts are
\usepackage{makeidx}         % allows index generation
\usepackage{graphicx}        % standard LaTeX graphics tool
\usepackage{multicol}        % used for the two-column index
\usepackage[bottom]{footmisc}% places footnotes at page bottom
\usepackage{hyperref}
\usepackage{siunitx}

\newcommand{\fixcomment}[1]{\textbf{\textit{[* #1 *]}}}

\makeindex             % used for the subject index
\begin{document}

\title*{Characterization of Pulsar Sources for X-ray Navigation}
\author{Paul S. Ray, Kent S. Wood, Michael T. Wolff}
\institute{Paul S. Ray \at Naval Research Laboratory, Washington, DC USA \email{paul.ray@nrl.navy.mil}
\and Kent S. Wood \at Praxis Inc., resident at Naval Research Laboratory, Washington, DC USA \email{kent.wood@tsc.com}
\and Michael T. Wolff \at Naval Research Laboratory, Washington, DC USA \email{michael.wolff@nrl.navy.mil}}
\maketitle

\abstract{The Station Explorer for X-ray Timing and Navigation Technology
(SEXTANT) is a technology demonstration enhancement to the Neutron Star Interior
Composition Explorer (NICER) mission, which is scheduled to launch in 2017
and will be hosted as an externally attached payload on the International Space
Station (ISS). During {NICER}'s $18$-month baseline science mission to
understand ultra-dense matter through observations of neutron stars in the soft
X-ray band, {SEXTANT} will, for the first-time, demonstrate real-time, on-board
X-ray pulsar navigation. Using NICER/SEXTANT as an example, we describe the factors
that determine the measurement errors on pulse times of arrival, including source
and background count rates, and pulse profile shapes. We then describe properties of the SEXTANT navigation pulsar catalog and prospects for growing it once NICER launches. Finally, we describe the factors affecting the prediction of pulse arrival times in advance, including variable interstellar propagation effect and red timing noise. Together, all of these factors determine how well a particular realization of an X-ray pulsar-based navigation system will perform.}

%\abstract{Each chapter should be preceded by an abstract (10--15 lines long) that summarizes the content. The abstract will appear \textit{online} at \url{www.SpringerLink.com} and be available with unrestricted access. This allows unregistered users to read the abstract as a teaser for the complete chapter. As a general rule the abstracts will not appear in the printed version of your book unless it is the style of your particular book or that of the series to which your book belongs.\newline\indent
%Please use the 'starred' version of the new Springer \texttt{abstract} command for typesetting the text of the online abstracts (cf. source file of this chapter template \texttt{abstract}) and include them with the source files of your manuscript. Use the plain \texttt{abstract} command if the abstract is also to appear in the printed version of the book.}

\section{Introduction}

As described in Chapter \fixcomment{Bernhardt}, X-ray emitting pulsars can
be used as the basis of a spacecraft navigation system
\cite{SheikhPinesEtAl2006}.  Pulsars are neutron stars formed
in supernova explosions that are born with spin periods of $\sim 10$ ms.  
As they age they spin down to periods as long as several seconds under the 
influence of electromagnetic torque associated with a rotating dipolar magnetic field.  
The torque removes rotational kinetic energy,
and a fraction of this released energy loss manifests as beamed radiation
produced along open field lines emanating from the polar caps.  This 
emission, which can extend from radio frequencies to $\gamma$-rays,
appears as pulses repeating at the neutron star's
rotation period.

A pulsar of this type is a rotation-powered pulsar (RPP). There are
other types, including accreting neutron stars and magnetars, but they
are of less interest for navigation, because of their transient behavior and 
complex period evolution.  The magnitude of rotational instabilities observed
in RPPs (known as timing noise) is positively
correlated with the spin period derivative \cite{antt94}. 
Younger, more energetic, pulsars have more
timing noise and this limits their navigational usefulness. The most promising
class of pulsars for navigation are millisecond pulsars (MSPs). These
pulsars have been spun up, gaining
angular momentum through accretion of material from binary companions for prolonged periods,
of order 10$^8$ years.  During this phase, the torque associated
with accretion causes them to reach spin periods in the range 1--10
ms, faster than the spins they had at their formation.  
During that process, the magnetic field decreases from $\sim$ 10$^{12}$ G
to values near 10$^8$ G, which has the consequence
of reducing the electromagnetic torque and contributes to their being 
more stable clocks than young RPPs.   Such pulsars
were first discovered in radio observations but (as described in Chapter
\fixcomment{Wood}) in 1992
X-ray pulsations were discovered from some MSPs \cite{BeckerTrumper1993}. With their
smaller magnetic fields, MSPs have not only lower period derivatives but also less
timing noise.  As a consequence, MSPs have far longer lifetimes than RPPs.  

Overall, about a dozen of the known MSPs show X-ray pulses.  These
fall into two sub-classes.  In one sub-class pulses are faint, soft,
and broad but pulsars in this class are typically the best intrinsic clocks.  
In these cases the emission is thermal from a hot spot on
the stellar surface with an effective temperature of about 2 keV. The
other sub-class comprises MSPs that show bright, hard, narrow 
pulses and also exhibit a larger level of timing
noise, or even rare glitches as in the case of PSR B1821$-$24 \cite{cb04}.
Notably, over 75 percent of all MSPs are in binary systems.

\subsection{X-ray MSPs and Navigation}

In a pulsar navigation system, measurements of pulse
times of arrival (TOAs) are compared with predictions of those TOAs
based on a timing model and the current estimate of the state vector
(position and velocity) of the spacecraft. Measured differences between
the measured and predicted TOAs are processed by an extended
Kalman filter to derive an improved estimate of the spacecraft state
vector. The performance of such a navigation
system is determined by the precision of the TOA measurements and the
accuracy of the timing model. Unlike a man-made satellite navigation system, the
source location, brightness and modulation properties cannot be
designed into the system.  Yet, 
on long timescales (years) MSPs have noise characteristics comparable to atomic
clocks. The engineering challenge is to exploit the naturally-occurring 
MSPs to devise a practical navigation system.
The long term stability of MSPs means that such a system can be highly autonomous, 
needing only very infrequent updates of the pulsar timing model parameters.

%\begin{figure}
%\includegraphics[width=2.5in]{WhiteGeeseInSnowstorm.pdf}
%\caption{Measures of variance are compared for millisecond pulsars and 
%representative atomic clocks.  The metric for clocks is the Allan variance.  
%The measure for pulsars is designed to be comparable and is based upon
%third order polynomials fitted to phase differences.  This is needed because,
%unlike an atomic clock, a pulsar does not have a value for a true frequency 
%known \textit{a priori}.  \label{fig:PulsarSigma}}
%\end{figure}

In this paper, we describe the properties of the specific pulsar
sources and of the measurement system that together determine the
accuracy of each TOA measurement and the prospects for increasing the
catalog of useful pulsars. We also characterize the error
budget for the timing model. Understanding the sources of error will
allow for realistic expectations for the performance of pulsar-based
navigation systems.

\section{NICER/SEXTANT}
\label{sec:1}

While many points we will be making pertain to any pulsar navigation system, we
illustrate them using the particular example of the SEXTANT technology
demonstration that will  be carried out using the NICER payload on the
International Space Station (ISS).  The performance goals for SEXTANT give
an idea of what is feasible with current instruments on the ISS
platform.  The navigation accuracy requirement is 10 km in the worst
of three directions, these being conventionally radial, in-track, and
cross-track.  This implies measuring pulsar arrival times to a typical accuracy of 33
$\mu$s.  The stretch goal is 1 km, or 3 $\mu$s.  

NICER,is an X-ray timing
experiment whose design has been highly optimized for millisecond
pulsar observations \cite{SPIE:NICER_06_2014}. It comprises 56
identical X-ray telescopes, each consisting of a concentrating X-ray
optic and a single-pixel silicon drift detector (SDD).  The complete payload
has a mass of 263 kg and draws 337 W when in nominal operation.
The peak total collecting area is over 1700 cm$^2$, about twice that of
the XMM-Newton EPIC-pn camera, and it is sensitive to X-rays in the
0.2--12 keV band. A key feature for precise timing observations is
that each photon is time-tagged to an accuracy of 100 ns (1 $\sigma$)
by reference to an onboard GPS receiver.  NICER is scheduled to be
launched in early 2017 and will be attached to the ISS for a minimum
24 month mission.  Its primary science goals include constraining the
dense matter equation-of-state by modeling the energy-dependent X-ray
pulse profiles of millisecond pulsars, probing the rotational
stability of MSPs, and searching for new periodic and quasi-periodic
timing behavior from neutron star systems. It will also perform a wide
variety of other observations that will be driven by an open-access
guest observer program.   The NICER instrument has been made as large
as possible to provide high-quality science data.  As such it is not a
direct prototype of an avionics package for X-ray navigation, but 
results from navigation experiments can be scaled to packages
of other sizes in other situations. In particular, for an interplanetary
cruise where accelerations are small it would be feasible to do very long
integrations, trading time for collecting aperture.  This would permit 
avionics packages much smaller than NICER to be effective. 

The SEXTANT team has developed a flight software application (called XFSW)
that will run on the NICER flight processor
\cite{Mitchell:2014a,Mitchell:2015a}.  This application will enjoy
onboard access to the X-ray photon data stream collected by the NICER
X-ray timing instrument and can use that data stream to extract TOAs from
particular pulsars as they are observed.  The primary goal of the
software application is to
demonstrate, for the first time, the determination of the navigation
solution for a spacecraft (in this case, the ISS) based solely on
X-ray pulsar measurements onboard and in real time.  This will be done
during one or more intervals of two weeks each, during which the NICER
observation schedule will be a sequence of MSPs optimized for
navigation performance, rather than science.  However, because much of
the NICER science requires large amounts of observing time on MSPs, 
without imposing significant constraints on times when these data are collected, the
SEXTANT-optimized observations will be fully useful for astrophysics
as well.

\section{TOA Measurement Accuracy}
\label{sec:TOA}

A TOA is a pulse arrival time, computed from a set
of observed photon times recorded by a detector. The photon times are converted to pulse
phases using a predictive model of the pulse period and
are compared to a high signal-to-noise pulse template to determine the
best estimate for the arrival time of a fiducial point on the template
profile.  

For binned data, this can be computed by finding 
the lag that maximizes the cross-correlation between a binned profile 
and the template.  In fact, for binned data (as one might get from radio 
measurements or from very
high photon count rates), a better method is to do the lag
determination in the Fourier domain, taking advantage of the Fourier
Shift Theorem, as described by \cite{Taylor1992}.  

\begin{figure}
\includegraphics[width=4.5in]{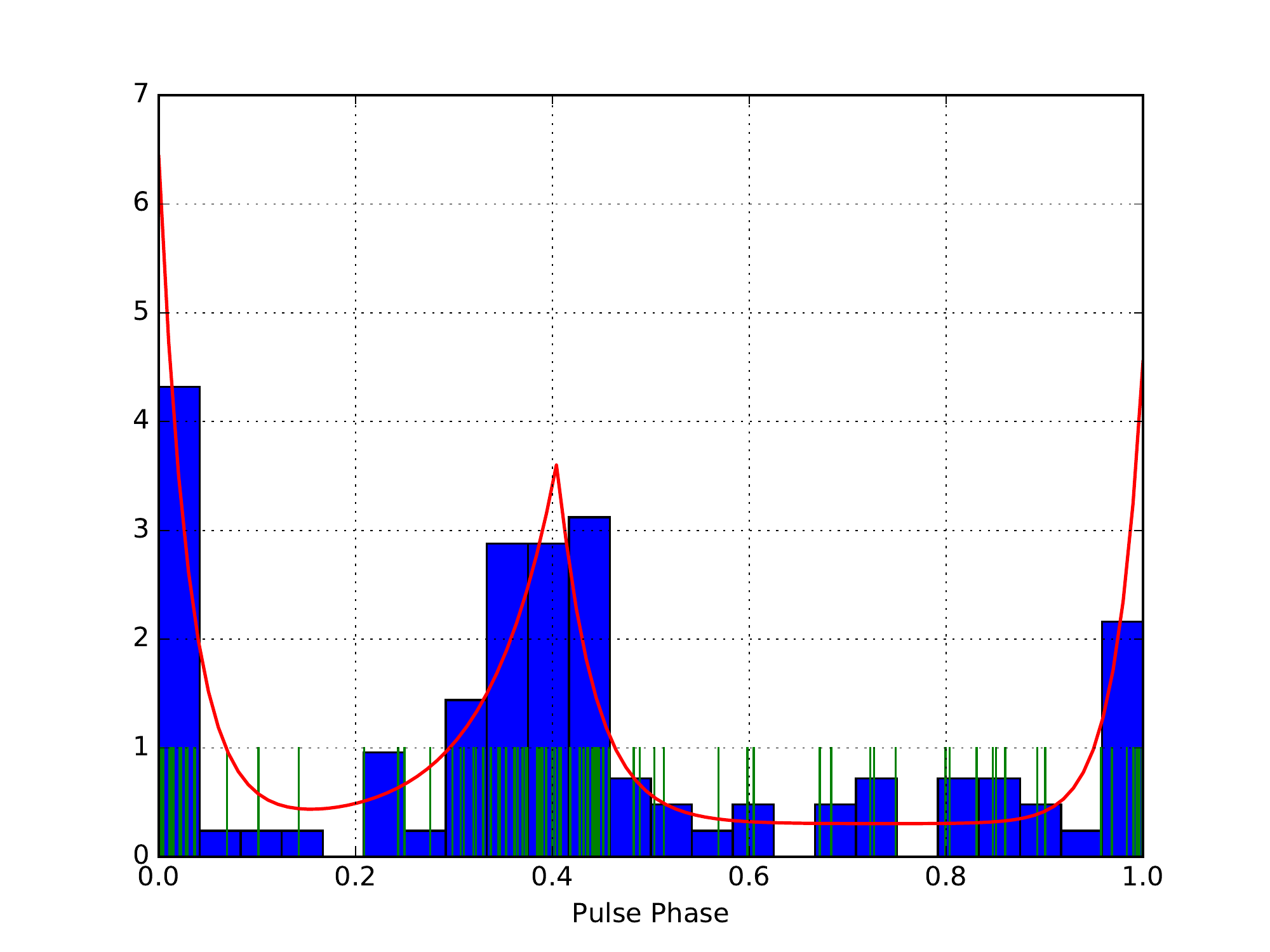}
\caption{The pulsar template (shown in red) approximates the ideal waveform that would
be obtained in an extremely long integration.  Actual integrations will have small 
numbers of photons (shown as green tick marks). Event times can be processed directly,
or binned (blue histogram) for computational speed at some loss in time resolution.   
\label{fig:PulsarTemplateWithCountsOverlaid}}
\end{figure}

For low-rate photon
data, information is lost in the binning process and maximum
likelihood methods are preferred \cite{rkp+11,Livingstone2009,XNAVieee:2015}.  Figure
\ref{fig:PulsarTemplateWithCountsOverlaid} illustrates the situation for low photon rates. 

In either case,
an algorithm running on the onboard processor determines
the time of arrival, or TOA, of the pulse at the detector.  In this section
we describe the
various contributions to the error budget for the TOA.

The first contribution to the TOA accuracy is the accuracy of the time
stamp that provides the reference time for the TOA. As all
measurements will be biased by any error in this time standard, this
time must be accurate to better than the accuracy goal. For SEXTANT,
the 100 ns accuracy provided by the NICER GPS time stamps is more than
sufficient for even the stretch goal.
Because time is referenced to GPS, for SEXTANT the pulsar measurements 
are used only for
determining the position (and velocity) of the ISS. A future pulsar
navigation system should be independent of GPS. This will require a
stable local clock onboard, which the pulsar measurements would then
steer to a pulsar-based time standard (akin to how GPS receivers steer
their local oscillators to match GPS time). This will require more
pulsars providing measurements to account for the additional dimension
of the problem (a minimum 4 pulsars, rather than 3). It will also
set a stability requirement on the onboard clock. For example, we
could set a requirement on the clock of $<$\SI{3}{\us} drift over an
interval long enough for at least 4 measurements to be processed
(perhaps 10$^3$ seconds). This corresponds to a fractional frequency stability of
$3\times 10^{-9}$. To provide robustness in an actual system the
required clock performance would probably need to be substantially
better than this. It is worth noting that if the system is scaled down
to a smaller detector then the interval required to collect
measurements at a particular accuracy will grow and the requirements
on the onboard clock stability will get more stringent.

Even with perfect time stamps, the accuracy of the arrival time will
be determined by the noise statistics of the measured signal. The arrival
times of the photons can be modeled as a non-homogeneous Poisson
process (NHPP) with a rate function represented by the pulse template
profile, designated by $h(\xi)$, where $\xi$ is pulse phase in the range $[0,1)$ and $h$ is normalized such that $\int_0^1 h(\xi) d\xi = 1$. 
For a fully specified statistical
model, the limiting performance of any estimator is set by the
Cram\'{e}r-Rao Lower Bound (CRLB), which yields the  limiting or 
best accuracy in phase determination obtainable in a particular situation 
\cite{GolshanSheikh2007,XNAVieee:2015}.

Golshan \& Sheikh (2007, eq. 24), derive the Cram\'{e}r-Rao lower bound on the accuracy of an
arrival time measurement ($\tau$)  as:
\begin{equation}
\sigma_\tau > \frac{P}{\sqrt{I_p T}},
\end{equation}
where $P$ is the pulse period, $T$ is the integration time, and 
\begin{equation}
I_p = \int_0^1 \frac{(\alpha h'(\xi))^2}{\beta + \alpha h(\xi)} d\xi
\end{equation}

The lower bound optimally exploits the pulse shape
information in the light curve, in terms of $h(\xi)$ and  
its derivative $h'(\xi)$, and converts to measured results using
coefficients $\alpha$ and $\beta$ that characterize amplitudes of the signal
and background in the detector system, respectively.  

The values of $I_p$ for the SEXTANT pulsars are listed in Table \ref{tab:catalog}.  These 
are numerically integrated from the template pulse profiles. In many cases,
it is not convenient to compute these integrals and analytic approximations
may be useful in certain limits. These also can assist in visualizing the 
scalings with the parameters. In the table below, we show the expressions
for $I_p$ in four cases, for purely gaussian 
or sinusoidal pulse profiles, and in the
zero background ($\beta = 0$) case and the low signal case. Here, we define $\rho = \frac{\alpha}{\beta \sigma \sqrt{2 \pi}}$. For the Gaussian low signal case, the expression is valid for $\rho<1$ and the number of terms that need to be evaluated in the series gets smaller as $\rho$ approaches 0.

\begin{center}
\setlength{\tabcolsep}{10pt}
\setlength\extrarowheight{10pt}
\begin{tabular}{llll}
   & \multicolumn{1}{c}{$h(\xi)$} & $I_p (\beta = 0)$ & \multicolumn{1}{c}{$I_p (\beta \gg \alpha)$} \\
   \hline
Gaussian & $\frac{1}{\sigma\sqrt{2\pi}} e^{-(\xi-\xi_0)^2/2\sigma^2}$ & $\alpha/\sigma^2$ & 
$\frac{\alpha}{\sigma^2} \rho \sum_{n=0}^{\infty} (n+2)^{-3/2} (-1)^n \rho^n$
 \\
Sinusoid & $1 - \cos(\xi)$ & $(2 \pi)^2 \alpha$ & $\frac{\pi^2\alpha^2}{2(\alpha+\beta)}$ \\
\end{tabular}
\end{center}

The CRLB improves with cumulative
integration time, $T$, as 1/$\sqrt{T}$ for as long as Poisson statistics
dominate the limit.  In SEXTANT practice, it is not anticipated that one
will integrate long enough for other error sources such as time stamp
accuracy to become dominant.  A rough rule of thumb for estimating the CRLB intuitively
is that it equates to a characteristic width of the pulse divided by the signal-to-noise (SNR)
expressed in standard deviations.

\begin{figure}
\includegraphics[width=4.5in]{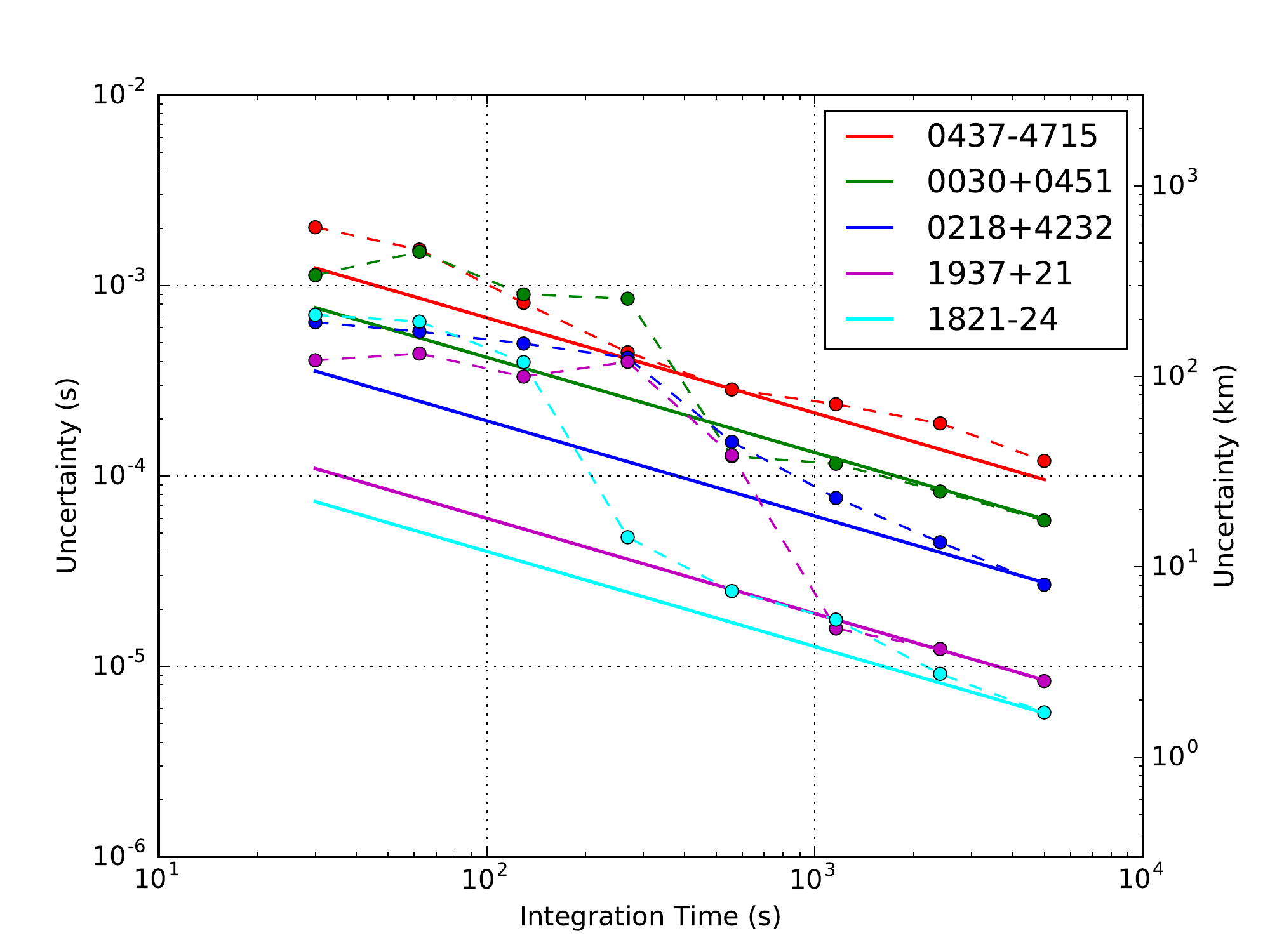}
\caption{Comparison of TOA accuracies (data points and dashed lines), based on event-by-event 
simulations and the actual SEXTANT measurement processing algorithm, for several SEXTANT 
target pulsars with the Cram\'{e}r-Rao Lower Bound (CRLB, solid lines).  
At large integration times the measurement uncertainties approach the CRLB. 
\label{fig:CRLBfor3PSRs}}
\end{figure}

Figure \ref{fig:CRLBfor3PSRs} shows the simulated measurement uncertainty for
several SEXTANT target pulsars vs. integration time, as compared to the CRLB. At long integration times, the measurement uncertainties approach the CRLB, as expected.

We note that the CRLB above is appropriate for situations where a single parameter is being determined (the pulse phase). If the uncertainty in the state vector is large enough and the orbit sufficiently dynamic (such as for the ISS in low Earth orbit), the measurement system may need to fit for multiple parameters (e.g. pulse phase and frequency, corresponding to a Doppler shift). In this case the CRLB on the TOA uncertainty is larger by a factor of two \cite{XNAVieee:2015}.

%As a concrete example consider a pulse 
%that is a pure Gaussian with standard deviation $\sigma$, further supposing there is
%zero background ($\beta$ =0) .  
%Then,  the CRLB evaluates to  $\sigma/(\alpha T)$  giving a standard deviation
%$\sigma / \sqrt{N}$, where N is
%the number of photons collected.  In this example the characteristic duration, $D$, is $\sigma$
%and$ \sqrt{N}$ is the SNR.  
%
%The rule of thumb for estimating CRLB ``by eye'' must be applied consistently, \textit{e.g.}, if one 
%chooses to characterize pulse width by an effective $\sigma$ then comparison is made 
%with $\sqrt{Variance}$ as given by the CRLB formula, but if ithe waveform is characterized 
%instead by its full-width half-maximum the same calculated result
%is multiplied by 2.33. More generally, D depends on the 
%exact pulsar waveform and the SNR depends on both $\alpha$ and 
%$\beta$,  as ${\alpha / \sqrt(\alpha + \beta)}$.  
 
The CRLB depends on several factors including astrophysical issues intrinsic to
the sources and particulars of the observing system. These factors influence the source selection and drive the optimal design of the instrument.  We now examine them one by one.

\subsection{Source count rate and its Estimation for Simulations} 

\begin{figure}
\includegraphics[width=4.0in]{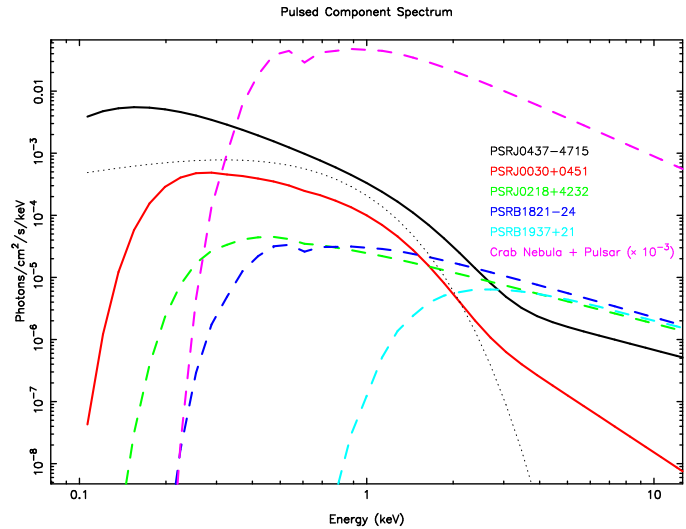}
\caption{X-ray spectra of several pulsars potentially useful for
navigation applications. The thin dotted line is a 0.2 keV blackbody with no 
interstellar absorption, for comparison.  Note that the magenta line is the 
Crab nebula and pulsar divided by 1000 to get it onto the same scale. \label{fig:mspspec}}
\end{figure}

The measured count rate from a source depends on both 
the properties of the source (its flux and spectrum) and the detector
system (its collecting area and quantum efficiency, as a function of
energy). The pulsed X-ray emission from MSPs can be dominated either
by thermal emission from the surface or by non-thermal emission from
the magnetosphere \cite{Zavlin2007}. In addition, the low-energy flux from the
source can be absorbed by differing amounts of material along the line
of sight.  See Figure \ref{fig:mspspec} for a comparison of several different MSP
spectra. Thermal spectra with minimal absorption produce
counts mostly in the 0.1--1 keV band, while non-thermal spectra,
particularly those with significant absorbing columns, demand
sensitivity extending out to several keV. For NICER, the effective
area curve is included in the online
WebPIMMS\footnote{\url{http://heasarc.gsfc.nasa.gov/cgi-bin/Tools/w3pimms/w3pimms.pl}} tool so count rates can be
easily estimated for any specified source flux and spectrum.  It
should also be noted that once NICER is in orbit, it will become the best
available instrument for determining both the X-ray spectra and pulsar
light curves, and that the latter may turn out to vary with energy
because of the varying blend of thermal and non-thermal components at
different energies.  This means that the initial models and templates
used in SEXTANT may be refined as NICER progresses and replaced with
versions of higher fidelity.

\subsection{Background count rate}

Unpulsed backgrounds from a variety of mechanisms contribute Poisson noise and reduce the accuracy of any TOA measurement.

A portion of the background is determined by the field of view of the
instrument and the performance of the imaging or concentrating
system. Any flux hitting the detector that comes from the diffuse
X-ray background or any neighboring sources simply contributes to the
background.  These contributions can be reduced by a smaller field of
view, at the expense of requiring a more precise instrument pointing
system. Similarly, any unpulsed flux from the pulsar source itself
provides an irreducible background. For SEXTANT the diffuse X-ray
background in the 3 arcmin FOV yields a background contribution of
0.15 counts/s.

In addition, particles or high energy photons that make it to the
detector from other than the optical path can interact and create
spurious events that contribute to the background.  For NICER, this rate
varies as the ISS moves in its orbit and encounters different trapped
particle environments, as well as with solar activity. Another
significant contribution to the radiation background for NICER comes
from the Soyuz spacecraft docked at the ISS, whose $\gamma$-ray
altimeters contain powerful radioactive sources. Together, these
radiation backgrounds are estimated to contribute 0.05
counts/s to the NICER background rate.

Because the spectral shapes of most background components are different from 
the target pulsars,
the performance can be optimized by making an energy selection that
minimizes the CRLB for a source. In addition, if there are nearby
sources making a significant background contribution, the instrument
pointing can be offset to reduce that contribution.  While the offset 
pointing reduces the signal from the source, in some cases it may
reduce the background from the nearby contaminating source in a way that more than compensates.

\subsection{Pulse Shapes}

\begin{figure}
\includegraphics[width=4.0in]{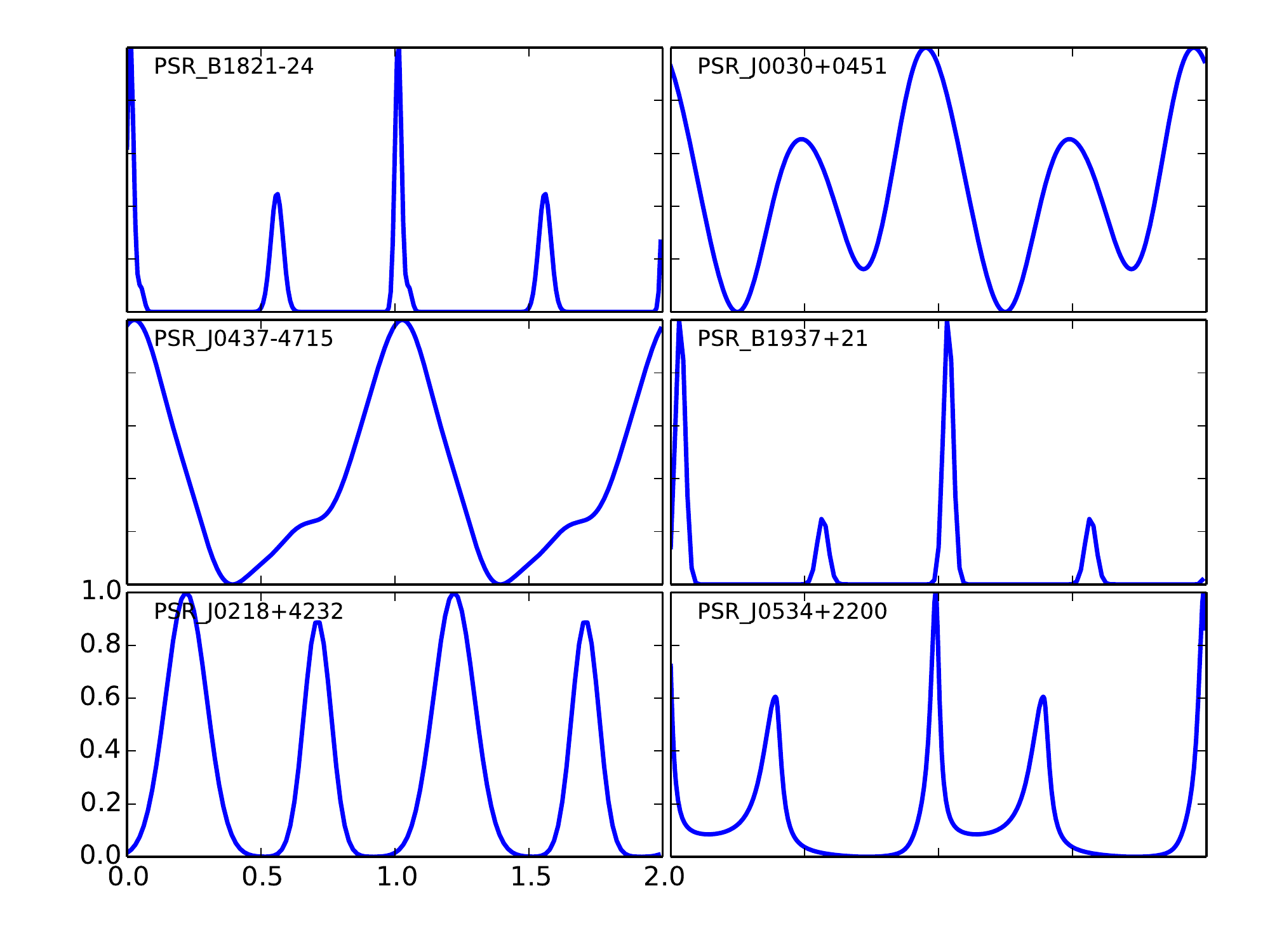}
\caption{Pulse profile templates for 6 of the SEXTANT target pulsars, showing the wide range
in light curve shapes.  For each pulsar two pulse cycles are displayed. \label{fig:templates}}
\end{figure}

Finally, the CRLB is determined by the pulse shapes of the pulsars.
All things being equal, pulse profiles with narrow features and steep gradients perform better
than those with smooth, nearly-sinusoidal, profiles. However, it was noted 
earlier that there is an anti-correlation between sharp pulses and 
overall clock stability for actual MSPs.  The pulse shape
templates for six pulsars in the SEXTANT catalog are shown in Figure \ref{fig:templates}.

\section{Source Catalog}

The selection of sources for the SEXTANT catalog is driven by the performance requirements.
SEXTANT's minimum performance requirement is being able to determine
the ISS orbit to 10 km RMS (worst direction) by the end of a
two-week experiment.  The stretch performance goal is 1 km RMS. Since
the speed of light is 300 m/\SI{}{\us}, these location accuracies
correspond to timing accuracies of \SI{30}{\us} and \SI{3}{\us},
respectively. TOA measurements (or model predictions) that are
substantially worse than this contribute little to meeting the SEXTANT
goals, although they may be useful in some circumstances such as in 
cold-start situations, or to resolve integer phase ambiguity issues
with the primary MSPs. This provides a criterion for selecting
individual sources to be included in the navigation catalog. However,
the properties of the catalog as a whole also need to be considered
when assessing system performance.

\begin{table}
\caption{SEXTANT Navigation Pulsar Catalog\label{tab:catalog}}
\setlength{\tabcolsep}{12pt}
\begin{tabular}{lrrrrrr}
\hline\hline
PSR & P (ms) & \multicolumn{1}{c}{$\alpha$ (c/s)} & \multicolumn{1}{c}{$\beta$ (c/s)}  & $I_p$ & $\sigma_\mathrm{CRLB}$ (\SI{}{\us}) \\
\hline
Crab Pulsar & 33.00 & 660.0 & 13860.2  & 56841.6 & 3.3  \\
B1937+21 & 1.56  & 0.029 & 0.24  & 23.3& 7.1 \\
B1821$-$24 & 3.05 & 0.093 & 0.22 & 240.5 & 4.7\\
J0218+4232 & 2.32 & 0.082 & 0.20 & 5.6 & 22.9  \\
J0030+0451 & 4.87 & 0.193 & 0.20 & 5.4 & 49.3 \\
J1012+5307 & 5.26 & 0.046 & 0.20  & 0.5 & 168.6 \\
J0437$-$4715 & 5.76 & 0.283 & 0.62 & 2.9 & 79.7 \\
B1509$-$58   & 151.25 & 6.5 & 0.20 & 347.8 & 191.2 \\
\hline
\end{tabular}
\end{table}

Table \ref{tab:catalog} shows the primary catalog of navigation sources currently
being used for SEXTANT. The pulse period, source ($\alpha$) and background ($\beta$) 
counting rates expected in NICER, $I_p$, and the CRLB estimate for the TOA uncertainty in
an 1800 s measurement are included. 
Note that the Crab pulsar is much brighter than the others and has a high $\alpha$ but
also a high $\beta$ that mostly comes from the off-pulse emision of the pulsar and its surrounding
nebula.  For most of the other sources $\beta$ is about 0.2 c/s, but J0437$-$4715 has
a higher rate because of the contaminating influence of a nearby source.  

\begin{figure}
\includegraphics[width=4.0in]{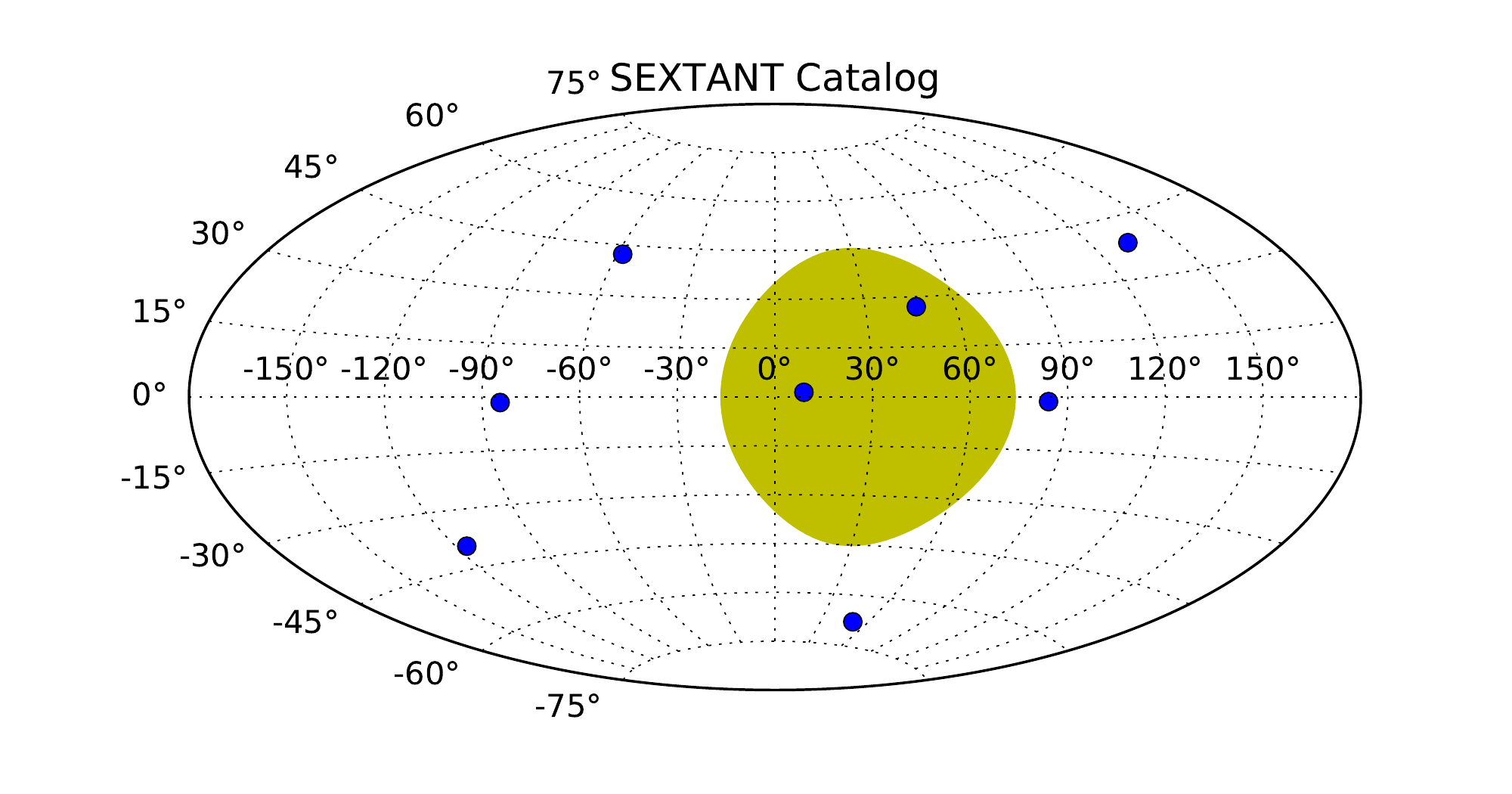}
\caption{Sky map in ecliptic coordinates, showing the pulsars in the SEXTANT 
catalog. The shaded region represents the solar exclusion region (shown here at one instant 
in time), which moves horizontally on the map over the course of a year.
\label{fig:skymap}}
\end{figure}

Figure \ref{fig:skymap} shows the distribution on the sky of the pulsars used by
SEXTANT.  The pulsars do not
concentrate preferentially along the Galactic plane because of the large scale height
of the millisecond pulsar population and the fact that most X-ray detected MSPs
are relatively nearby.  Thus they are a nearly
isotropic population, providing a good geometry for navigation.

Source visibility constraints limit when each source can be observed.
There are exclusions for Sun, Moon, Earth blockage, and ISS structure
blockage. In the figure, the solar exclusion zone is a moving
circle centered on the horizontal axis, which is the Ecliptic plane.

The source catalog in Table \ref{tab:catalog} may also change as new pulsars are
discovered and characterized. Many new
candidates are being found using the \textit{Fermi Gamma-ray Space
Telescope} and in follow-ups with ground-based radio observations.
In addition, NICER itself is expected to detect new X-ray pulsations
from several millisecond pulsars as part of its science program.
As these pulsars are discovered and characterized they are assessed
for their navigational potential.

\section{Model Accuracy}

Any error in the prediction of when a pulse should arrive directly translates
into an error in the derived navigation solution. The timing models used to generate
the predictions must be maintained by a program of radio, X-ray, or $\gamma$-ray
observations, informed by the best
current understanding of astrophysical mechanisms at
work in the sources.  Typically the highest precision TOAs are obtained from
radio observations, but care must be taken to account for variable propagation effects
that affect radio observations, but not X-ray observations (which are effectively
at infinite frequency). 

A useful source of timing information comes from pulsar timing array programs, designed to detect gravitational waves. (In principle,
gravitational waves make additional minute contributions to the phase
errors in received pulses.) Currently the NANOGrav project \cite{NANOGrav2013} is 
timing several of the SEXTANT pulsars, and the Nan\c{c}ay observatory in France 
has a long-term timing program on several others (some overlapping), importantly 
including the source B1821$-$24. The SEXTANT team has signed Memoranda of 
Understanding with both of those groups, providing for sharing of the timing
data for X-ray navigation purposes.
In addition, several of the pulsars are observed continuously in
$\gamma$-rays by $\textit{Fermi}$ \cite{kerr15}. Jodrell Bank observes the Crab
pulsar daily\footnote{\url{http://www.jb.man.ac.uk/pulsar/crab.html}} and the Parkes Pulsar Timing Array (PPTA) monitors 20 pulsar, including the important PSR J0437$-$4715 \cite{PPTA16}.
After launch, NICER itself will be an excellent source of precision timing 
information for the SEXTANT pulsars.

These data are used to derive parameterized timing models that predict the arrival
of any individual pulse to high precision, typically using a pulsar timing code like
\textsc{Tempo2} \cite{hobbs_tempo2_2006}.  Many of these parameters are 
deterministic, meaning that they are a measurement of a physical property 
of the system (e.g. spin frequency, position, proper motion, binary system parameters)
that can be used to predict TOAs forward in time with 
uncertainty bounded by the uncertainty in the determination of the model parameters.
However, there are some effects that are not stationary, or are stochastic.  When
these are present, the accuracy of the predictions will degrade with time, regardless
of the precision with which the model parameters are determined from the current data.
Fortuitously, these effects are rather small for MSPs. We discuss some of them below.

\subsection{DM variations}

\begin{figure}
\includegraphics[width=4.5in]{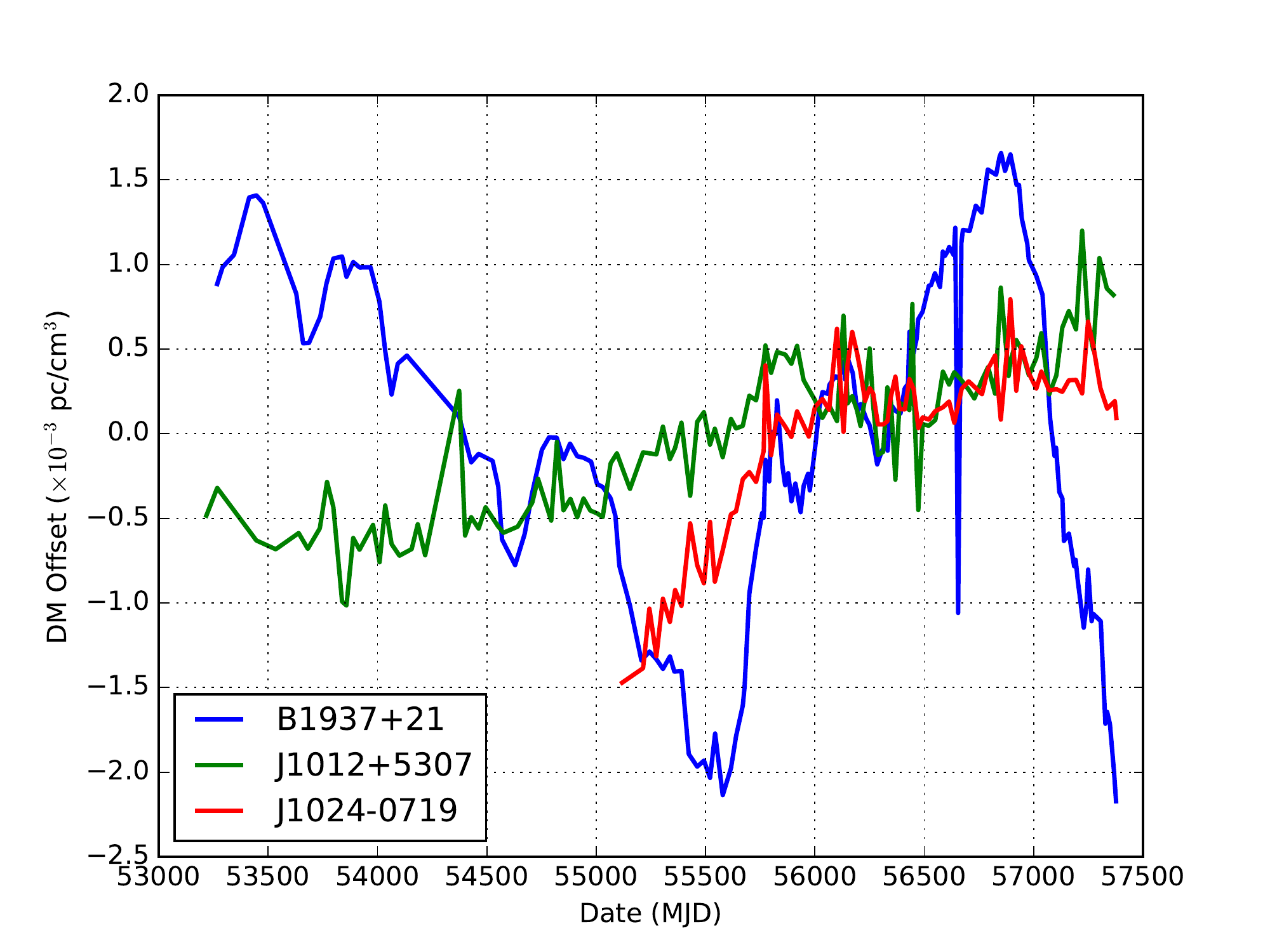}
\caption{Dispersion Measure (DM) variations can produce errors
in extrapolation from radio data to the X-ray band. These data are from NANOGrav timing measurements \cite{NANOGrav9yr}. For reference, the time offset caused by a DM error of 
$1.0 \times 10^{-3}$ pc cm$^{-3}$ from 1400 MHz to infinite frequency is 2 $\mu$s.
\label{fig:DMVariations}}
\end{figure}

The dispersion measure (DM) characterizes how the radio pulse is
delayed as a function of frequency caused by the frequency dependence
of the group velocity of a radio wave traveling in the tenuous plasma of
interstellar space.  The time delay relative to a signal of infinite frequency
is \cite{Handbook}
\begin{equation}
\Delta t = \mathcal{D} \times \frac{DM}{f^2}
\end{equation}
where $\mathcal{D} = 4148.8$ MHz$^2$ pc$^{-1}$ cm$^3$ s, and $f$ is the observing
frequency.  If the DM were constant there
would be a constant correction between the radio pulse arrival time
(at a particular frequency) and that in X-rays.  In principle this delay can
be measured and removed by making timing observations at two or more radio frequencies.
However, variations in the
column density of free electrons as the line of sight moves through the complex and turbulent interstellar medium cause stochastic wandering of the DM that is not
predictable over long time scales. The DM variation versus time for several SEXTANT
pulsars is shown in Figure \ref{fig:DMVariations}.

\subsection{Red noise and model extrapolation}

In addition to unpredictable DM variations, some MSPs exhibit red noise akin
to the timing noise seen in young RPPs.  This is modeled as a stochastic process
with a steep spectrum. As an example, in an analysis of 9 years of NANOGrav timing
data, red noise was detected in 10 out of 37 MSPs. This red noise causes the arrival 
times to drift relative to the deterministic model by a few microseconds over 
timescales of a few years.

Care must be taken when extrapolating timing models for pulsars with significant red noise contributions.  If the red noise is modeled using a high order polynomial, it will
extrapolate very poorly and the predictions will diverge from the true arrival times very quickly. A better method is to construct an optimal extrapolation based on the
statistical properties (i.e. the covariance matrix) of the red noise \cite{dch+12}. 
This method is implemented in the \texttt{interpolate} plugin for \textsc{Tempo2}.

The Crab pulsar is an important special case.  Unlike the other
SEXTANT sources it is a young, bright RPP and not a MSP.  It is
much brighter than the MSPs, so can be detected in a far shorter observation, 
yet has much more timing
noise.  Over short intervals and with good support from ground
observations it can be an important contributor to navigation performance, particularly in a highly dynamic situation (such as an orbital maneuver).
To maintain prediction accuracies of 10 $\mu$s, ephemeris updates are 
needed roughly every three days when the Crab is being employed for navigation.  
Figure 10 shows period excursions for the Crab over an interval of a month and
also illustrates the impracticality of trying to extrapolate outside
the observed interval.

For MSPs, the parameters are known well enough, and the effects of red noise and 
DM variations are small enough, that a timing model can predict phase to the accuracy
required for SEXTANT for months into the future.

\begin{figure}
\includegraphics[width=4.5in]{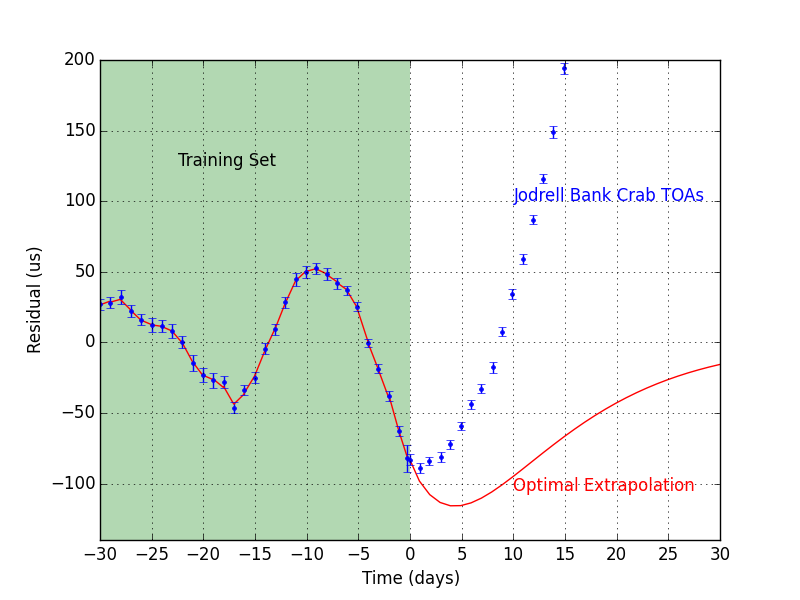}
\caption{The growth of residuals with days elapsed since the end of measurements, for
the Crab pulsar.  The model was fit to the data in the green region, and the red curve shows the optimal extrapolation into the future. The actual arrival times (blue points) diverge from the extrapolation rather quickly, so
for navigational purposes and the Crab timing model must be updated frequently throughout the 
interval of use.  MSPs, however, are much less noisy and for them extrapolations
beyond the fitted interval are viable for far longer. 
\label{fig:CrabExtrap}}
\end{figure}

\section{Conclusion}

\label{sec:5}

Pulsar navigation has been discussed for some years as a natural analog to global navigation satellite systems (GNSS) that may have important applications in cases of GNSS unavailability or for missions that travel deep into interplanetary or
interstellar space \cite{patent}.

As we have shown, the performance of such a system depends on the properties of the pulsars and measurement system, as well as the ability of the pulsar timing models to predict
pulse arrival times into the future.

The SEXTANT demonstration using the NICER X-ray timing instrument will be a major 
milestone in the development of X-ray navigation. The SEXTANT results will inform
the development of realistic navigation instruments based on X-ray pulsar observations. In addition, NICER should discover several new X-ray millisecond
pulsars that can contribute to navigation applications.  

\begin{acknowledgement}
SEXTANT work at NRL is supported by NASA NPR NNG13LM70I. 
\end{acknowledgement}

\bibliographystyle{spphys}
\bibliography{Ray}
\end{document}